\documentclass[a4paper]{article}
\usepackage[ansinew]{inputenc}
\usepackage{times}
\usepackage[T1]{fontenc}
\usepackage{graphicx}
\usepackage{geometry}
\usepackage{amssymb}
\usepackage{amsmath}

\usepackage{rotating}

\usepackage{xcolor}
\colorlet{shadecolor}{gray!25}
\usepackage{subfig}
\usepackage{float}

\author{Ludwig A. Hothorn,\\ 
Im Grund 12, D-31867 Lauenau, Germany\\ (e-mail: ludwig()hothorn.de. Retired from Leibniz University Hannover)}

\title{Claiming trend in toxicological and pharmacological dose-response studies:\\ an overview on statistical methods and related R-Software}

\begin{document}

\maketitle
\begin{abstract}
There are very different statistical methods for demonstrating a trend in pharmacological experiments. Here, the focus is on sparse models with only one parameter to be estimated and interpreted: the increase in the regression model and the difference to control in the contrast model. This provides both p-values and confidence intervals for an appropriate effect size. A combined test consisting of the Tukey regression approach and the multiple contrast test according to Williams is recommended, which can be generalized to the generalized linear (mixed effect) model. Thus numerous variable types occurring in pharmacology/toxicology can be adequately evaluated. Software is available through CRAN packages. The most significant limitation of this approach is for designs with very small sample sizes, often in pharmacology/toxicology.
\end{abstract}

\section*{The problem}
One of the main statistical principles in pharmacology/toxicology is:  \textit{'Characterization of Dose-response Relationships Inferred by Statistically Significant Trend Tests'} \cite{Kodell1991}. \\
But the vast majority of articles on trend tests deal with time trends. Exactly this is not discussed here, as the specific dependency between the times, e.g. 2014, 2015 and 2016, is modeled. Trends in dose-response dependencies, recommended in guidelines, e.g. for in-vivo micronucleus assay \cite{OECD474}, assume independent doses in a randomized design (classification I). The dose is thus a grouped variable, e.g. 0, 10, 50 mg/kg, whereby the dose levels can be considered either as qualitative levels of a factor (analysis of variance model) or quantitative values of a covariate (regression model) (or both; see below) (classification II). The essential classification depends on the objective, and this can be very different: simply to show an increasing global trend, to estimate a specific dose, such as MED, NOAEL (in the sense of Paracelsus) or using a model-based approach to estimate a lower confidence interval for a regulatory defined dose, such as $ED_{50}$ or the benchmark dose (BMD) (classification III). A related overview on model-based approaches is available \cite{Ritz2015}.  Some authors \cite{Kodell2009} argue to avoid trend test-based approaches (NOAEL estimation) altogether because of their design dependency and to use the generally superior model-based approaches, e.g. the BMDL (where L stands for lower confidence limit). Because of the usually few doses, the small sample sizes, the non-interpolation, the critical model-assumption dependence and the uncertainty of setting limits in BMD model, I would not prefer a BMDL of a NOAEL on principle. \\ A misunderstanding between toxicologists and statisticians is the simple definition of what is actually a trend. Often trend tests are used which were formulated for linear trends, such as the Armitage trend test \cite{Armitage1955} and the Jonckheere trend test \cite{Jonckheere1954}. But a specific shape of dose-response dependence is rarely an a-priori assumption, but rather a result of the evaluation. Therefore, trend tests are needed which are sensitive to as many shapes as possible, as well as supra- and sub-linear shapes, e.g. the common in pharmacology/toxicology plateau-like shapes, and also allow a statement to be made as to which shape of the alternative is most likely. We describe trend tests for near-to-linear trend, any shape trend, strictly monotone trend and non-monotone trend, especially with down-turn effects at high doses (classification VI).\\
Nowadays, the existence of a trend and its magnitude is usually represented by a p-value (or a derived star).
Its brevity and simplicity are bought at the expense of numerous serious disadvantages. Therefore, the biologically relevant choice of an effect measure and its confidence interval is the better alternative. For this purpose, the slope in 2-parametric regression models or the difference/ratio to control is particularly suitable (classification V).\\ Numerous statistical tests are based on normally distributed homoscedastic error terms in a randomized one-way layout. But in pharmacology/toxicology, many other endpoints such as proportions, counts, time-to-events, etc. occur in more complex systems, including random terms such as within-litter dependencies. One needs trend tests in the generalized linear model (GLM) and generalized linear mixed effect model (GLMM). Also robust variants for any distribution and even informative variance heterogeneity, e.g. increasing with effect increase (classification VI).\\
A basic idea is a model that is as simple ('spares') as possible, which is realized by exactly one, interpretable parameter, e.g. the slope in regression models or the difference to the control in contrast tests. Of course you can find much better fitting models, e.g. polynomials, splines or 5-parametric nonlinear models. But both their comparison across endpoints, sex, species, studies, relevant in pharmacology/toxicology, and their biological interpretability are rather difficult (classification VII).\\
 Limiting factors are the very small sample size per dose group $n_i$ common in pharmacology/toxicology, up to the triplicates in the Ames assay, and the usually few dose levels (k=3+1): in terms of the false+/false- decision ratio, model discrimination, robustness, etc. In practice, asymptotic tests are used (which assume $n\rightarrow \inf$) and sometimes the bias is massive. There is no universal solution to this problem, but in some places the mitigation of this limitation is pointed out.\\

\fbox{
\parbox{0.9\linewidth}{
\large
Classification:  \small
 (bold ... discussed here; italic ... not here)
\begin{itemize}
	\item [I:] \textbf{independent dose groups in a randomized design} vs. \textit{dependent time points} 
	\item [II:] \textbf{a) dose modeled qualitatively, b) dose modeled quantitatively, c) jointly} 
	\item [III:] \textbf{a) just claiming global trend}, \textit{b) estimate NOAEL (MED), c) model-based estimation of ED50, BMDL} 
	\item [IV:] \textbf{tests for a) near-to-linear trend, b) any shape trend, c) strictly monotone trend and d) non-monotone trend} (especially with down-turn effects at high doses)
	\item [V:] avoid p-values. \textbf{Use an appropriate effect measure and its confidence interval for interpretation}
	\item [VI:] \textbf{a) trend tests in GLM, GLMM, b) robust trend tests}
	\item [VII:] \textbf{using sparse models with preferably only one interpretable parameter}
 \end{itemize}
}}
\normalsize

\section*{Motivating examples}
Three data sets were selected. One from regulatory toxicology (repeated administration of sodium dichromate dihydrate in a chronic study on rat), a second from environmental toxicology (aquatic assay using ceriodaphnia dubia) and a third from veterinary pharmacology using beagle dogs.
In the left panel of Figure \ref{fig:Boxplot} the number of total brood in ceriodaphnia dubia after treatment with increasing tank concentrations of nitrofen \cite{Bailer1994} are show, in the middle panel the blood urea nitrogen (BUN) endpoint of the 13-week study with sodium dichromate dihydrate \cite{NTP2002} where a downturn effect at high dose occur, and in the right panel are the 24 hours diuresis values for 0, 1, 2, and 4 mg/kg i.m. furosemide to treat congestive heart failure in dogs \cite{Bieth2016}. The occurring data diversity is thus somewhat depicted. Increasing and decreasing trends, monotone trends and those with downturn effects at high doses (and those with plateau effect), counts and continuous endpoints, variance homogeneity and heterogeneity, designs with $(k=3+1)$, $(k=4+1)$ and $(k=5+1)$:

\begin{figure}[htbp]
	\centering
		\includegraphics[width=0.286\textwidth]{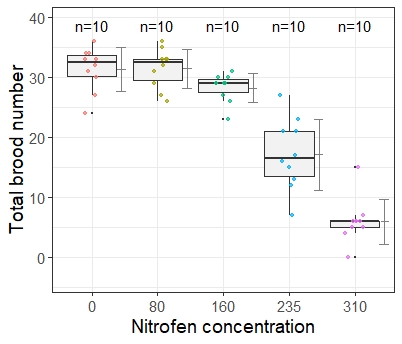}
		\includegraphics[width=0.286\textwidth]{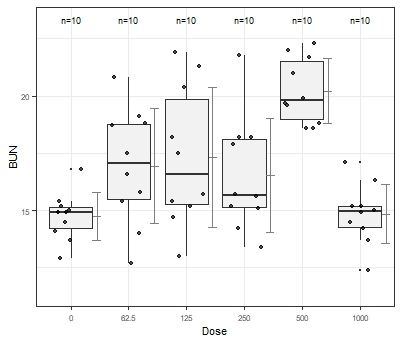}
		\includegraphics[width=0.286\textwidth]{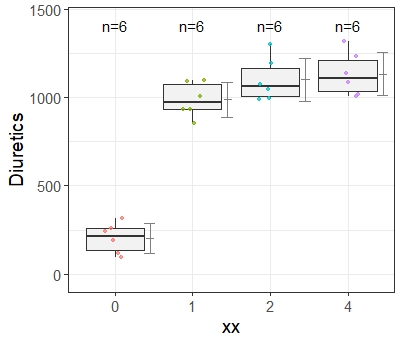}
	\caption{Boxplots for example data: nitrofen (left), blood urea nitrogen (middle), furosemide (right)} 
	\label{fig:Boxplot}
\end{figure}

\section*{Trend tests}
\subsection*{Trend tests where dose is modeled qualitatively: multiple contrast tests}
To consider the dose as a factor with qualitative levels seems absurd at first, since the dose levels are known. However, the difference between the applied dose and the concentration at the target and the assumptions-sparing modeling can be exactly that.\\
Due to the sparse model strategy (class. VII), only linear contrast tests are considered here.
A contrast is a suitable linear combination of means $\bar{x}_i$ (or other effect sizes, such as ratio-to-control, odds ratio, hazard ratio):
$	\sum_{i=0}^k c_i\bar{x}_i$, where the contrast coefficient  $c_i$ are appropriate chosen weights. Specific for all experiments in pharmacology/toxicology is the comparison to the control $i=0,...,k$. A single contrast test is its standardized version $t_{Single\: contrast}=\sum_{i=0}^k c_i\bar{x}_i/S\sqrt{\sum_i^k c_i^2/n_i}$ and $\sum_{i=0}^k c_i=0$ ensures a $t_{df,1-\alpha}$ distributed level-$\alpha$-test.
A multiple contrast test (MCT) is defined as maximum test is: $t_{MCT}=max(t_1,...,t_q)$ which follows jointly  $(t_1,\ldots,t_q)^\prime$ a $q$-variate $t$- distribution with degree of freedom $df$ and the correlation matrix  $R$  $\Rightarrow$ depending on $c_i, n_i$ but also $s_i, \rho_i,...$ and may be complex. The choice of the weights $c_i$ and hence the particular contrast matrix defines the special version, e.g. for a simple balanced design with k=2: 
Dunnett one-sided  test \cite{Dunnett1955}
        \begin{table}[htbp]
            \begin{tabular}{ c  c c r c c c c }
        $c_i$ & C & $T_1$ & $T_2$ \\
        $c_a$& -1 & 0  & 1   \\
        $c_b$& -1 & -1 & 0  \\
        \end{tabular}
        \end{table}
				
    Williams one-sided procedure (formulated as multiple contrast \cite{Bretz2006})
   \begin{table}[htbp]
            \begin{tabular}{ c  c c r c c c c }
         $c_i$ & C & $D_1$ & $D_2$ \\
        $c_a$ & -1 & 0   & 1   \\
        $c_b$ & -1 & 1/2 & 1/2  \\
        \end{tabular}
        \end{table}
The simple Williams contrast illustrates the idea and interoperability of an MCT: either the comparison of $D_2$ vs. C is significant (i.e. strict monotone), or to the pooled $(D_2+D_1)/2$, i.e. a plateau (or both, or neither ($H_0$).  Either multiplicity-adjusted p-values or simultaneous confidence limits (here one-side lower one) are available:
    $[\sum_{i=0}^k c_i\bar{x}_i - S* t_{q,df,R,2-sided,1-\alpha}\sqrt{\sum_i^k c_i^2/n_i}]$
This makes the Williams test the recommended trend test in pharmacology/toxicology: sensitive to some monotonic and partially non-monotonic forms, a comparison to the control, easily interpretable confidence intervals for the required effect sizes (difference of proportions \cite{Hothorn2010}, risk ratio or odds ratio of proportions \cite{Hothorn2016}, ratio-to-control estimates \cite{Hothorn2011}, relative effect sizes \cite{Konietschke2012}, hazard rates \cite{Herberich2012}, multiple endpoints \cite{Hasler2012}, heteroscedastic error terms \cite{Herberich2010}, poly-k-adjusted tumor rates \cite{Schaarschmidt2008}).\\
Sometimes the question arises whether to use the Dunnett-test (unrestricted $H_1$) or the Williams-test (restricted $H_1$)? The simple answer is use both \cite{Jaki2013}:
\begin{table}[htbp]
            \begin{tabular}{ c  c c r c c c c }
         $c_i$ & C & $D_1$ & $D_2$ \\
        $c_a$ & -1 & 0   & 1   \\
				$c_b$ & -1 & 1   & 0   \\
        $c_c$ & -1 & 1/2 & 1/2  \\
		\end{tabular}
        \end{table}		
The slight increase in conservatism (correlated contrasts!) is usually overcompensated by the simultaneous interpretability of disordered or ordered effects.\\

\hfill \break

\fbox{
\parbox{0.9\linewidth}{

\textbf{Considering the doses as qualitative factor levels, the William test can be recommended using the R-package \textit{multcomp}
}}}

\subsection*{Trend tests where dose is modeled quantitatively: Tukey-type regression models}
There are numerous approaches to the relationship between response and dose formulated as quantitative covariates, but because of the sparse-model strategy quasilinear regression models are considered here (Notice, The MCP-Mod approach, see below, can be seen as a linear regression with a single predictor variable \cite{Thomas2017}).  Tukey's trend test \cite{Tukey1985} is close to pharmacological principles and quite simple: the simultaneous considering of three regression models for the metameters $D_i$, (untransformed) $rank(D_i)$ (ordinal-transformed) and $log(D_i)$ (lin-logarithmic transformed dose). This joint test represents also max-T-test, where the joint distribution over these linear (generalized) models (instead of just contrasts) is achieved by the multiple marginal model approach (mmm) \cite{Pipper2012} where a related CRAN package is available \cite{Schaarschmidt2019}.

\subsubsection*{Trend tests where dose is modeled jointly qualitatively and quantitatively}
One can combine the advantages of both Williams (qualitative) and Tukey (quantitative) test by formulating a related joint max-T test \cite{Hothorn2020b}, since both models belong to the class of linear models and thus the common adjustment by means of mmm is available. This test has the average best balanced power, i.e. shows high power for either linear or plateau shapes. Of course it shows a reduced power, if one knew a-priori that there is a linear relationship and one would only use a linear regression model. But this is far from reality in pharmacology/toxicology, because the shape is the result of the bioassay, not an a-priori assumption. Because of the rather high correlated model, this decline is acceptable compared to the power gain and more precise interpretation of the likely shape alternative/model.\\
\hfill \break

\fbox{
\parbox{0.9\linewidth}{

\textbf{Considering the doses as quantitative covariate levels, the Tukey test can be recommended, even its joint test together with Williams-type contrasts using the R-package \textit{tukeytrend}.
}}}

\hfill \break

Analyzing the BUN example shows the appropriateness of the Tukey-Williams joint test: although the quantitative regression part does not show any monotonous increase, while at least a Williams contrast (most in the alternative the contrast $(1000+500)/2-0$) shows a trend up to the reversal point at 500 mg/kg. The effect size is the means value difference-to-control (and the regression slope in the regression part- not in the alternative) where the  most in the alternative effect size is $(\bar{y}^{BUN}_{1000}+ \bar{y}^{BUN}_{500})/2-\bar{y}^{BUN}_{0})=2.8mg/dl$ with a lower limit of 1.8 mg/dL.

\begin{table}[ht]
\centering
\footnotesize
\caption{Tukey-Williams Trend Test: BUN example} 
\label{tab:wuL}
\begin{tabular}{lrr}
  \hline
Model & Test statistics & $p$-value \\ 
  \hline
Tukey: arithmetic & -0.275 & 0.905 \\ 
Tukey: ordinal & 1.436 & 0.231 \\ 
Tukey: ari-logarithmic & 1.436 & 0.231 \\ 
Williams: $1000-0$ & 0.219 & 0.760 \\ 
Williams: $(1000+500)/2-0$ & 6.525 & $<0.0001$ \\ 
Williams: $(1000+500+250)/3-0$& 5.517 & $<0.0001$ \\ 
Williams: $(1000+500+250+125)/4-0$ & 5.472 & $<0.0001$ \\ 
Williams: $(1000+500+250+125+62.5)/5-0$ & 5.565 & $<0.0001$ \\ 
   \hline
\end{tabular}
\end{table}

The R-Code is quite simple using the package \textit{tukeytrend}: first a simple linear model is fitted into the object \textit{mod1}. Its estimators are used in the function \textit{tukeytrendfit} to perform the MaxT test using the 3 regression models (ari, ord, arilog) and the William contrasts (\textit{cytpe=Williams}), where  \textit{ddf} uses the empirical degree of freedom and \textit{vcov} the sandwich variance estimator to be robust to heteroscedasticity. In the object \textit{ex1} the function \textit{glht()} (from the package \textit{multcomp}) estimates the adjusted p-values, or even better the simultaneous confidence limits for a potential monotonous increase of the BUN values with dose (see the data and the executable R code in the Appendix).   \\
A further underrated issue is the trend test when heteroscedasticity occur- this can lead to a bias. As long as $n_i$ are not too small, the use of a sandwich variance estimator is recommended \cite{Herberich2010} (see the BUN example above).

\hfill \break
\footnotesize
\begin{verbatim}
mod1 <- lm(BUN ~ dose, data=mm)
tw1 <- tukeytrendfit(mod1, dose="dose", ddf="residual",
                     scaling=c("ari", "ord", "arilog", "treat"),
                     ctype="Williams")\\
ex1<-summary(glht(tw1 mmm, tw1 mlf,vcov = sandwich, alternative="greater"))
\end{verbatim}
\normalsize
\subsection*{An underrated fact: shape specificity}
The shape of the dose-response dependence is the result of the experiment, rarely an a-priori assumption. In addition to the existence of a global trend, the information on what shape of the dose-response dependence exists is relevant once again. This is quite challenging.\\
Nevertheless, trend tests that are designed for a linear alternative, such as the CA test for proportions \cite{Armitage1955} or the nonparametric Jonckheere-test \cite{Jonckheere1954}, are widely used. This is disturbing, because one rarely wants to test for linear trend exactly, and a loss of power occurs, e.g. if there is a plateau at high doses which is not uncommon in pharmacology/toxicology. In the sense of an average high power for any monotone shape, trend tests which are sensitive to several alternatives are to be preferred. These are e.g. the Williams and Tukey tests demonstrated above whereby their joint test is best in pharmacology/toxicology from this point of view, sensitive to a wide range of alternatives, with still acceptable conservatism \cite{Schaarschmidt2020}.\\
Notice, for some issues in toxicology it is unacceptable to infer global trend with the usual trend tests, although there is a significant downturn at high doses. One connects trend with monotonous, and a downturn effect at high dose(s) is just not monotonous. Here you can use a combination of trend test and test between control and high dose: as an intersection-union tests \cite{Lin2019}, an union-intersection test with the complete power approach \cite{Hothorn2020b} or surprisingly just the simple pairwise (or contrast) test control vs. high dose.\\
On the other hand, there are situations, like the muta-tox issue in in-vitro mutagenicity assay, which, despite a possible down-turn at high doses, would still like to decide on global trend. The toxicologist would simply follow a trend test up to the optical maximum effect (and ignore the higher doses).  Statistically correct this can be achieved by means of maxT-test over all possible maximum doses (ignoring the higher ones) \cite{Bretz2003}.

\subsection*{Trend tests in generalized linear models (GLM)}
Another contradiction exists: most trend tests assume normally distributed, homoscedastic errors, but most endpoints in pharmacology/toxicology are 
skewed distributed or are counts, proportions, time-to-event variables \cite{Szocs2015}. One solution is the use of the GLM, where suitable link-functions are used to transform the different endpoint types appropriately. This is a general but asymptotic approach, i.e. it requires a very large sample size and can cause problems with the common small $n_i$.

\textbf{Proportions} are quite common in pharmacology/toxicology, such as microscopic incidences or crude mortality/ tumor rates. For both the Williams-trend test \cite{Hothorn2010} and the Tukey test \cite{Hothorn2020b} related GLM-modifications exist, whereas the add-1 correction for common-used small $n_i$ can be recommended to control the $\alpha$-level approximately \cite{Schaarschmidt2008a}. Note two very different models. On the one hand, the occurrence of a condition such as a tumor is recorded at the level of the experimental unit, e.g. the animal. On the other hand, proportions are repeatedly recorded within an experimental unit, such as (e.g., the number of
micronucleated erythocytes per scored polychromatic erythrocytes within a single mouse 
in the in vivo micronucleus assay \cite{Hothorn2009}. In the latter case, a link function for overdispersion can be used, since pooling the individual proportions within the experimental unit usually underestimates the variance.\\
Similarly, \textbf{counts} can be evaluated, such as number of micronuclei; also with or without modeling of overdispersion \cite{Hothorn2016}.  Count data are particularly sensitive to the underlying assumption, e.g. for ties and variance heterogeneity structure.
The CRAN package \textit{cotram} \cite{Siegfried2020} offers flexible count transformation models where count responses may arise from various and complex data-generating processes. Because the endpoint are counts with some tied data, the Dunnett test assuming approximate normal distribution is not appropriate here. Instead the cotram package \cite{Siegfried2020} offers count transformation models, providing a simple but flexible approach for the regression analysis of count responses arising from various, and possibly complex, data-generating processes. (Notice, also for continuous variables there are robust methods with the most likely transformation model behind \cite{Hothorn2019, Hothorn2018}). \\
 The above example with the total brood count in an aquatic assay on ceriodaphnia dubia with nitrofen \cite{Bailer1994} is used to demonstrate this approach:

\footnotesize
\begin{verbatim}
data("nitrofen", package="boot") 
nitrofen$Conc<-as.factor(nitrofen$conc)
library(multcomp)
library(cotram)
NSH<-cotram(total~Conc, data=nitrofen,order=5)
WISH<-glht(NSH, linfct = mcp(Conc = "Williams"), alternative="less")
\end{verbatim}

\begin{table}[ht]
\centering
\footnotesize
\begin{tabular}{lrrrr}
  \hline
 Contrast & Estimate & SE & Test statistics & p-value \\ 
  \hline
$310-0$ & -12.9 & 2.32 & -5.56 & $<0.0001$ \\ 
 $(310+235)/2-0$ & -10.0 & 1.75 & -5.75 & $<0.0001$ \\ 
 $(310+235+160)/3-0$ & -7.2 & 1.27 & -5.68 & $<0.0001$ \\ 
 $(310+235+160+80)/4-0$ & -5.4 & 1.03 & -5.28 & $<0.0001$ \\ 
   \hline
\end{tabular}
\caption{Williams-type p-values for count transformation model} 
\end{table}
\normalsize

All Williams contrasts are seriously in the alternative, most is the contrast $(310+235)/2-0$ with the effect size of an odds ratio, i.e. a significant decreasing trend exists.
Notice, the count transformation model belongs not directly to a GLM (here a negative binomial link function can be used). Attention should be paid to the specificity of data, where near-to-zero counts occur frequently, especially in control. Simple transformations may be appropriate here \cite{Jaki2014}.\\

\textbf{Multinomial vectors} occur also, such as the differential blood count. There analysis is more complex and related multiple contrast modifications are available \cite{Schaarschmidt2017}.\\
\textbf{Time-to-event data} occur commonly, such as time-to-death or time-to-tumor occurrence in long-term carcinogenicity assays. These survival functions can be compared by the Williams-trend test assuming a Cox proportional hazard model \cite{Herberich2012}.\\
Because of their high toxicological relevance, the evaluation of \textbf{graded histopathological findings} (with severity scores) is particularly interesting. Very different models can be helpful, especially if near-to-zero categories occur in the control: collapsing categories \cite{Green2014} (or a closed permutation test \cite{Hothorn2016}, Chapter 2.1.8.1), cumulative link model for ordinal endpoints {cite{Hothorn2016}, Freeman-Tukey transformation models or nonparametric contrast tests using the relative effect size estimator (see the section below).\\
Due to the complex interaction between tumor development and mortality, the evaluation of crude tumor rates can be biased. Therefore, a mortality-adjusted analysis is indicated, relatively simple on the basis of \textbf{poly-k estimators}, which do not need the cause of death. Individual weights $w_{ij}=\left( t_{ij}/t_{max} \right)^k$ considering the individual mortality pattern are used where $t_{ij}$ is time of death of animal $j$ in dose $i$ and the tuning parameter $k$ is data-dependent. The weights result in adjusted sample sizes $n_{i}^{*}=\sum_{j=1}^{n_{i}}w_{ij}$ to be used instead of the randomized number of animals $n_{i}$. Therefore adjusted proportions $p_{i}^{*}=y_{i} /n_{i}^{*}$ are used in the GLM instead of the crude tumor proportions $p_{i}=y_{i} /n_{i}$. This procedure is available for the Williams test \cite{Schaarschmidt2008a} and the Tukey test \cite{Hothorn2020b}.\\
As a somewhat extreme example of simultaneous inference, the tumor and mortality data of the US-NTP study on the carcinogenic potential of methyleugenol for the incidence of skin fibromas used \cite{NTP2000} (available in the CRAN package \verb|MCPAN|). 
Since dose either quantitative (Tukey) or qualitative (Williams) fits best, or the tuning parameter k=3 or k=6, we use a max(maxT)-test on these $((3+3)*2)$ marginal models in weighted GLM's for an identity link for difference of weighted proportions to control:

\footnotesize
\begin{verbatim}
me$weightpoly3[wt0] <- (me$death[wt0]/max(me$death))^3
me$weightpoly6[wt0] <- (me$death[wt0]/max(me$death))^6
fitpoly3 <- glm(tumour ~ dose, data=me, family=binomial(link="identity"),
                weight=weightpoly3)
fitpoly6 <- glm(tumour ~ dose, data=me, family=binomial(link="identity"),
                weight=weightpoly6)
library(tukeytrend)
ttpoly3 <- tukeytrendfit(fitpoly3, dose="dose",
           scaling=c("ari", "ord", "arilog", "treat"), ctype="Williams")
ttpoly6 <- tukeytrendfit(fitpoly6, dose="dose",
           scaling=c("ari", "ord", "arilog", "treat"), ctype="Williams")
compttpoly6 <- glht(model=ttpoly6$mmm, linfct=ttpoly6$mlf, alternative="greater")
tt36 <- combtt(ttpoly3,ttpoly6)
TT36 <- summary(asglht(tt36, alternative="greater"))
\end{verbatim}
\normalsize
Most in the alternative is the plateau-type Williams contrast for poly-6 adjustment, where as the difference to poly-3 is tiny in this particular example.\\

\begin{table}[ht]
\centering
\footnotesize
\caption{Tukey-Williams-type trend test models for poly k=3 and k=6 in the skin fibroma tumor data} 
\label{tab:exa14}
\begin{tabular}{llrr}
  \hline
Poly-k&Model & Test statistic & p-value \\ 
  \hline
3&Tukey: arithmetic & 1.77 & 0.0877 \\ 
 & Tukey: ordinal & 2.50 & 0.0181 \\ 
  &Tukey: ari-logarithmic & 2.48 & 0.0190 \\ 
  &Williams $150-0$& 1.91 & 0.0670 \\ 
  &Williams $(150+75)/2-0$ & 3.03 & 0.0041 \\ 
  &Williams $(150+75+37)/3-0$ & 3.83 & 0.0003 \\ 
  6&Tukey: arithmetic  & 2.32 & 0.0277 \\ 
  &Tukey: ordinal  & 2.98 & 0.0048 \\ 
  &Tukey: ari-logarithmic  & 2.96 & 0.0051 \\ 
  &Williams $150-0$ & 2.08 & 0.0471 \\ 
  &Williams $(150+75)/2-0$ & 3.18 & 0.0025 \\ 
  &Williams $(150+75+37)/3-0$ & 3.97 & 0.0002 \\ 
   \hline
\end{tabular}
\end{table}

\subsubsection*{Trend tests when adjusting against covariate and secondary factors is required}
Sometimes the consideration of a covariate, such as initial body weight, or an additional factor, such as sex, is important. Name-giving tests, such as the CA test, are usually difficult to extend - in GLM this is very easy \cite{Hothorn2016}. A typical application is the evaluation of organ weights, where the body weight can be modeled as covariates.

\subsubsection*{Trend tests in generalized linear mixed effect models (GLMM)}
The heterogeneity between repeated times, between litter mates, between cages, between tanks in aquatox-assays, the cells within the slides within the organ-specific samples of the Comet assay, or independent repeated studies on a particular test substance can be modeled as a random factor and thus the hierarchy of the different sources of variance can be reproduced correctly by GLMM. Such models are already complex and data-dependent, sometimes numerically unstable, especially when only a few levels can be considered. \\
Repeated bioassays are common for relevant chemicals by several sponsors, e.g. the glyphosate bioassays. Here, simplifying
the crude proportions of four studies (abbreviated by A-D) for malign lymphoma in male mice are considered \cite{Tarazona2017}.
\begin{figure}[htbp]
	\centering
		\includegraphics[width=0.450\textwidth]{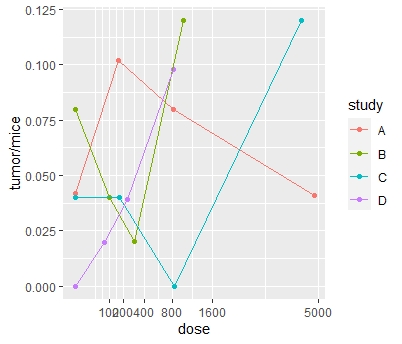}
	\caption{Crude malign lymphoma tumor rates of 4 carcinogenicity assays on glyphosate in male mice}
	\label{fig:Gly}
\end{figure}
Figure \ref{fig:Gly} shows considerable heterogeneity between the studies, not only with regard to the shape of the dose-response relationship, but also with regard to the spontaneous rates, especially with regard to the very different doses chosen - across orders of magnitude. This alone makes the qualitative modeling of the dose (Williams test) less appropriate. Therefore, the regression approach according to Tukey is the method of choice here, whereby the heterogeneity between studies with a GLMM is estimated. The function \textit{glmmPQL()} provides a robust penalized quasi-likelihood approach in GLMM (with the canonical link function \textit{binomial()}) and the internal function \textit{lmer2lm()} transforms GLMM-objects to GLM objects within the multiple marginal model approach \cite{Pipper2012}- the basis for the decisive function \textit{glht(mmm())} with in the package \textit{multcomp}.

\footnotesize
\begin{verbatim}
library(tukeytrend)
Lmic <- dosescalett(Lmice, dose="dose", scaling=c("ari", "ord", "arilog")) data
glmmari <- glmmPQL(fixed=cbind(tumor,mice-tumor) ~ doseari, random = ~ 1 +doseari |study,
                   family = binomial, data=Lmic)
glmmord <- glmmPQL(fixed=cbind(tumor,mice-tumor) ~ doseord, random = ~ 1 +doseari |study,
                   family = binomial, data=Lmic, niter = 100)
glmmarilog <- glmmPQL(cbind(tumor,mice-tumor) ~ dosearilog, random = ~ 1 +doseari |study,
                      family = binomial, data=Lmic)
lmari <- tukeytrend:::lmer2lm(glmmari)
lmord <- tukeytrend:::lmer2lm(glmmord)
lmarilog <- tukeytrend:::lmer2lm(glmmarilog)
linf <- matrix(c(0,1), ncol=2)
ttglmm <- glht(mmm("mari"=lmari, "mord"=lmord, "marilog"=lmarilog),
               mlf("mari"=linf, "mord"=linf, "marilog"=linf), alternative="greater")
\end{verbatim}
\normalsize

\begin{table}[ht]
\centering
\footnotesize
\caption{Odds ratios and their lower confidence limits using the Tukey trend test on crude malign lymphoma tumor rates of 4 carcinogenicity assays} 
\label{tab:exaGLY}
\begin{tabular}{lrr}
  \hline
Model & Odds ratio & Lower confidence limit \\ 
  \hline
Tukey: arithmetic & 1.000 & 0.999 \\ 
Tukey: ordinal & 1.071 & 0.976 \\ 
Tukey: ari-logarithmic & 1.214 & 0.960 \\ 
   \hline
\end{tabular}
\end{table}

For none of the 3 regression models is the lower limit greater than 1, i.e. there is no increasing trend, which is not surprising given these pronounced heterogeneities.

\hfill \break

\fbox{
\parbox{0.9\linewidth}{
\textbf{
Use the modifications of the Tukey-Williams joint test for non-normal endpoints and/or complex designs in the generalized linear (mixed effects) model.}
}}

\subsection*{Nonparametric trend tests}
Since the normal distribution assumption cannot be tested because of the small $n_i$, nor can it be derived from historical controls, nonparametric trend tests are often used, especially the Jonckheere-test and the global-ranking version of the Williams test \cite{Shirley1977}. For tied values (see above the graded findings), and for the usual heterogeneous variances, these tests are problematic. As an alternative, non-parametric tests for relative effect size can be used, which are also available for multiple contrast tests and thus as a modification of the Williams test \cite{Konietschke2012}. A corresponding CRAN package nparcomp \cite{Konietschke2015} is easily used for this purpose.

\subsection*{Effect measures and their confidence interval}
If one follows the recommendation to avoid p-values for the interpretation of trend tests, and instead to use effect size estimates adapted to the toxicological problem and its intervals (confidence intervals or rather compatibility intervals \cite{Hothorn2020c}), then the choice of a suitable effect size estimate is of primary importance. This contradicts the dominant choice of the difference-to-control as an effect size estimator. But there are much more effect sizes. One of them has to do with the scale of the response variable, e.g. the risk ratio for proportions or hazard rate for time-to-event variables. On the other hand, the effect size can significantly define the modeling objective, such as ratio-to-control or odds ratio for continuous outcome logistic regression \cite{Hothorn2019}.

\subsubsection*{Using ratio-to-control instead of the common-used difference-to-control}
Difference-to-control is the mostly used effect size, mostly hidden, because their confidence intervals are not explicitly interpreted, but simply the p-values. Sometimes one wants to interpret ratio-to-control, such as the k-fold rule of the Ames assay. The usual procedure of log transformation of the response variable can be problematic \cite{Schaarschmidt2013}. Instead, a direct modification of the Williams test is available for normally distributed, even heteroscedastic errors \cite{Hothorn2011} (numerically easily available in the package \textit{mratios}, \cite{Dilba2007}). In most cases the interpretation of a percentage change in pharmacology/toxicology is appropriate, but not for very small near-to-zero values in the control.

\hfill \break
\fbox{
\parbox{0.9\linewidth}{
\textbf{
Select effect sizes that are biologically appropriate and interpretable.}
}}

\section*{Further trend test issues}
Phase IIb randomized clinical dose-finding trials (RCT) are similar to the above assays, with the rather fewer doses (k=2, 3) arguing for low-parameter models \cite{Saha2019}. The common-used MCP-Mod approach represents a hybrid method consisting of contrast tests and related nonlinear models \cite{Pinheiro2014}.\\
Sometimes, after detecting a trend, one wants to determine a specific experimental dose, such as minimal effective dose (MED) in RCT or  no-observed adverse event level (NOEAL) (in toxicology). In doing so, one should pay attention to a possible bias due to the pooling properties of the trend tests.\\
We had only assumed randomized dose groups, but sometimes a trend was observed in continuous exposure, e.g. transforming into groups via post-hoc categorization which may be misleading \cite{Greenland1995}. \\
Sometimes you want to show that no trend exists. This cannot be done simply by a non-significant trend test, since  \textit{'absence of evidence is no evidence of absence'} \cite{Altman1995}. Instead, one tests the slope parameter for non-inferiority, whereby the determination of the threshold limit is a delicate problem.

\section*{Conclusions}
Trend tests sensitive to many possible shapes, considering dose qualitatively or quantitatively (or best jointly), 
formulating for a single, interpretable parameter, in the same (or similar)  framework for as many effect sizes (and its confidence limits) and thus variable types as possible are available nowadays- a methodical step forward. Not only the statistical methods are available, but also the corresponding software in form of CRAN packages. A structure like a construction kit was described above, whereby real data assays were evaluated exemplarily (further examples see \cite{Hothorn2016, Hothorn2020c, Hothorn2020a, Hothorn2020b}).\\

The biggest limitation are the widely used small to very small $n_i$, especially with often associated unclear data conditions. Sometimes in these situations the use of simple transformations, such as arcsin or FT, followed by the standard trend tests even makes sense - as least bad test.

\footnotesize
\bibliographystyle{plain}

\normalsize

\section*{Appendix: Data and R-Code}
\scriptsize
\begin{verbatim}
############################# BUN example
mm <-
  structure(list(BUN = c(15, 15.2, 14.9, 15.4, 16.8, 14.1, 14.9,
                         14.5, 13.7, 12.9, 15.8, 16.6, 12.7, 15.4, 14, 20.8, 19.1, 18.7,
                         18.8, 17.5, 21.9, 17.5, 21.3, 20.4, 18.2, 15.4, 15.2, 13, 14.7,
                         15.7, 15.7, 13.4, 15.1, 14.2, 15.2, 17.9, 18.2, 21.8, 18.2, 15.6,
                         22.3, 18.6, 22, 19.9, 19.6, 21, 21.7, 19.7, 18.8, 18.6, 14.9,
                         15.2, 15.2, 12.4, 13.7, 14.2, 15, 14.5, 16.3, 17.1),
                 Dose = structure(c(1L,1L, 1L, 1L, 1L, 1L, 1L, 1L, 1L, 1L,
                                    2L, 2L, 2L, 2L, 2L, 2L, 2L,2L, 2L, 2L,
                                    3L, 3L, 3L, 3L, 3L, 3L, 3L, 3L, 3L, 3L,
                                    4L, 4L, 4L, 4L, 4L, 4L, 4L, 4L, 4L, 4L,
                                    5L, 5L, 5L, 5L, 5L, 5L, 5L, 5L, 5L, 5L,
                                    6L, 6L, 6L, 6L, 6L, 6L, 6L, 6L, 6L, 6L),
                                  .Label = c("0", "62.5", "125", "250", "500", "1000"),
                                  class = "factor")), row.names = c(NA,-60L), class = "data.frame")


library("toxbox")
boxclust(data=mm, outcome="BUN", treatment="Dose", cluster=NULL,
         ylabel="BUN", xlabel="Dose", option="uni", psize=1.1,nsize=2.5,
         hjitter=0.3, legpos="none", printN="FALSE", white=TRUE, titlesize=8, labelsize=6)
###########################
library(tukeytrend)
library(multcomp)
library(sandwich)
mm$dose<-as.numeric(as.character(mm$Dose))
mod1 <- lm(BUN ~ dose, data=mm)
tw1 <- tukeytrendfit(mod1, dose="dose", ddf="residual",
                     scaling=c("ari", "ord", "arilog", "treat"),
                     ctype="Williams")
ex1<-summary(glht(tw1$mmm, tw1$mlf,vcov = sandwich, alternative="greater"))
T1<-fortify(summary(ex1))[, c(1,5,6)]
colnames(T1)<-c("Model","Test stats", "$p$-value")
library(xtable)
print(xtable(T1, digits=3, caption="Tukey-Williams Trend Test", label="tab:wuL"),
      include.rownames=FALSE, caption.placement = "top", sanitize.text.function = function(x){x})

############################### nitrofen example
data("nitrofen", package="boot") # total brood
nitrofen
library(toxbox)
boxclust(data=nitrofen, outcome="total", treatment="conc", xlabel="Nitrofen concentration",
         ylabel="Total brood number", psize=1.15, hjitter=0.15, vlines="fg", white=TRUE, printN=FALSE, legpos="none")

nitrofen$Conc<-as.factor(nitrofen$conc)
library(multcomp)
library(cotram)
NSH<-cotram(total~Conc, data=nitrofen,order=5)
WISH<-glht(NSH, linfct = mcp(Conc = "Williams"), alternative="less")
library(xtable)
library(ggplot2)
wish<-fortify(summary(WISH))
print(xtable(wish, caption="Williams-type p-values for count transformation model",
             caption.placement = "top", digits=4))
						
########### poly3- poly-6 example
library(MCPAN)
data("methyl", package="MCPAN")
data(methyl)
me <- methyl
# death: time of death, max(death)=length of study = 730 days
# poly-3 adjustment: (time of death/max(time))^k, k=3,
#for those animals without tumour at time of death!
# Compute the poly-3 (-k)- weights at the level of single animals
me$weightpoly3 <- 1
me$weightpoly6 <- 1
# Animals without tumour at time of death get corrected sample size
wt0 <- which(me$tumour == 0)
me$weightpoly3[wt0] <- (me$death[wt0]/max(me$death))^3
me$weightpoly6[wt0] <- (me$death[wt0]/max(me$death))^6
me$dosegroup <- me$group
levels(me$dosegroup) <- c("0", "37", "75", "150")
me$dose <- as.numeric(as.character(me$dosegroup))
# Notice, the  Modell use an identity link for difference to control,
fitpoly3 <- glm(tumour ~ dose, data=me, family=binomial(link="identity"),
                weight=weightpoly3)
fitpoly6 <- glm(tumour ~ dose, data=me, family=binomial(link="identity"),
                weight=weightpoly6)
library(tukeytrend)
ttpoly3 <- tukeytrendfit(fitpoly3, dose="dose",
                         scaling=c("ari", "ord", "arilog", "treat"), ctype="Williams")
ttpoly6 <- tukeytrendfit(fitpoly6, dose="dose",
                         scaling=c("ari", "ord", "arilog", "treat"), ctype="Williams")
compttpoly6 <- glht(model=ttpoly6$mmm, linfct=ttpoly6$mlf, alternative="greater")
tt36 <- combtt(ttpoly3,ttpoly6)
TT36 <- summary(asglht(tt36, alternative="greater"))
library("ggplot2")
library(xtable)
polyk36<-fortify(TT36) [, c(1,5,6)]
colnames(polyk36)<-c("Comparisons","Test stats", "p-value")
print(xtable(polyk36,  digits=c(0,0,2,4), caption="Tukey-Williams-type trend test models for poly k=3 and k=6", label="tab:exa14"),caption.placement = "top", include.rownames=FALSE)

####################################repeated glyphosat example
Lmice <- data.frame(
  study=c(rep("A",4), rep("B",4),rep("C",4), rep("D",4)),
  dose = c(0, 157, 814, 4841, 0,100,300,1000,
           0, 165, 838, 4348, 0,71,234, 810),
  Wscore=c("C","D1","D2","D2", "C","D1","D2","D2", "C","D1","D2","D2", "C","D1","D2","D3"),
  tumor = c(2,5,4,2, 4,2,1,6,2,2,0,6,0,1,2,5),
  mice = c(48,49,50,49,50,50,50,50, 50,50,50,50,51,51,51,51))
library(xtable)
LMI<-xtable(Lmice, digits=c(0,1,0,0,0,0))
print(LMI, include.rownames=FALSE,
      caption.placement = "top",size="\\footnotesize")
library(ggplot2)
  ggplot(Lmice, aes(y=tumor/mice, x=dose))  + geom_point(aes(color=study)) +
  geom_line(aes(color=study, group=study)) +
  scale_x_continuous(trans="sqrt",breaks=c(0,100,200,400,800,1600,5000))
library(tukeytrend)
Lmic <- dosescalett(Lmice, dose="dose", scaling=c("ari", "ord", "arilog"))$data

glmmari <- glmmPQL(fixed=cbind(tumor,mice-tumor) ~ doseari, random = ~ 1 +doseari |study,
                   family = binomial, data=Lmic)
glmmord <- glmmPQL(fixed=cbind(tumor,mice-tumor) ~ doseord, random = ~ 1 +doseari |study,
                   family = binomial, data=Lmic, niter = 100)
glmmarilog <- glmmPQL(cbind(tumor,mice-tumor) ~ dosearilog, random = ~ 1 +doseari |study,
                      family = binomial, data=Lmic)
lmari <- tukeytrend:::lmer2lm(glmmari)
lmord <- tukeytrend:::lmer2lm(glmmord)
lmarilog <- tukeytrend:::lmer2lm(glmmarilog)
linf <- matrix(c(0,1), ncol=2)
ttglmm <- glht(mmm("mari"=lmari, "mord"=lmord, "marilog"=lmarilog),
               mlf("mari"=linf, "mord"=linf, "marilog"=linf), alternative="greater")
l2<-summary(ttglmm)
CIG<-confint(ttglmm)
cigg<-CIG$confint

\end{verbatim}

\end{document}